\documentclass[aps,prl,showpacs,twocolumn,superscriptaddress,showpacs,groupedaddress,nofootinbib]{revtex4-1}  

\usepackage{color}
\usepackage{graphicx}  
\usepackage{dcolumn}   
\usepackage{bm}        
\usepackage{amssymb}   
\usepackage{amsmath,amssymb,amsthm,mathrsfs,amsfonts,dsfont} 
\usepackage[colorlinks=true,citecolor=magenta, linkcolor=darkblue]{hyperref}
\usepackage[dvipsnames]{xcolor}
\definecolor{darkblue}{rgb}{0.0, 0.0, 0.55}

\pdfoutput=1

\newcommand{\myhref}[1]{\href{https://arxiv.org/abs/#1}{{\color{darkblue} #1}}}

\def\be{\begin{eqnarray}}
\def\ee{\end{eqnarray}}
\def \bea {\begin{eqnarray}}
\def \eea {\end{eqnarray}}
\newcommand{\oBox}{\mathring{\Box}}

\newcommand{\oR}{\mathring{R}}

\def\id{\mathds{1}}
\def\K{\kappa}
\def \nn {\nonumber}
\def \la {\langle}
\def \ra {\rangle}

\def \m {\mu}
\def \n {\nu}

\def\Vol{\operatorname{Vol}}

\def\frac#1#2{{#1\over #2}}

\def\tr{\operatorname{tr}}

\begin{document}


\title{
Displacement Operators 
and Constraints on Boundary Central Charges
}

\author{Christopher Herzog$^{1}$, Kuo-Wei Huang$^{1,2}$ and Kristan Jensen$^{3}$\vspace{0.1cm}}
\affiliation{
$^{1}$C. N. Yang Institute for Theoretical Physics,
Stony Brook University,
Stony Brook,
NY 11794,
USA}
\affiliation{
$^{1,2}$Perimeter Institute for Theoretical Physics,
Waterloo, Ontario N2L 2Y5,
Canada,}
\affiliation{
$^{3}$Department of Physics and Astronomy,
San Francisco State University,
San Francisco, CA 94132,
USA}


\fontsize{10pt}{11.7pt}\selectfont

\begin{abstract}
Boundary conformal field theories have several additional terms
in the trace anomaly of the stress tensor associated purely with the boundary. 
We constrain the corresponding boundary central charges in three- and four-dimensional 
conformal field theories in terms of two- and three-point correlation functions
of the displacement operator.  
We provide a general derivation by comparing the trace anomaly with 
scale dependent contact terms in the correlation functions. 
We conjecture a relation between the $a$-type boundary charge in three dimensions  
and the stress tensor two-point function near the boundary.
We check our results for several free theories.

\end{abstract}
\maketitle

There is a strong argument for considering, from an abstract point of view, boundaries
in quantum field theory (QFT).
Boundary effects can be seen as a unifying theme in several areas where there has
been enormous progress in theoretical physics.
They are essential to understanding condensed matter
systems such as topological insulators and quantum impurity models.
D-branes, i.e.\ the boundaries of fundamental strings, gave us non-perturbative
insight into string theory and led to the second superstring revolution in the late 90s.  
In gauge-gravity duality, which provides windows both on
strongly-interacting quantum field theories and on quantum gravity, quantum 
fields fluctuate on the conformal boundary of anti-de Sitter space.
Entanglement entropy in field theory is usually defined with respect to spatial regions, 
introducing  an ``entangling'' surface which separates the regions.
Entanglement has given us new insight into renormalization group 
flow \cite{Casini:2004bw,Casini:2012ei,Casini:2017vbe}, and has deepened
our understanding of black hole thermodynamics \cite{Casini:2008cr} and 
energy conditions \cite{Balakrishnan:2017bjg}.  

Conformal field theories (CFTs) play a central role in QFT as fixed points of the renormalization group flow.  
It seems reasonable that boundary conformal field theories (bCFTs) should 
play a similarly central role in the study of QFT with a boundary.  
More specifically, given the important role of trace anomalies in CFT without a boundary, it is reasonable 
to expect that boundary terms in the trace anomaly should be important as well. 

We begin with a general discussion of the boundary terms in the trace anomaly including definitions
of the anomaly coefficients $a_{(3d)}$, $b$, $b_1$, and $b_2$.
We prove
that the coefficients $b$ and $b_2$ are related to two- and three-point functions 
of the displacement operator.
Our main results are \eqref{brel} and \eqref{b1rel}.  
We conjecture that the $a_{(3d)}$ coefficient satisfies a related constraint~\eqref{arel}, from 
which follows a lower bound~\eqref{bound3d} on $a_{(3d)}/b$.  
We then demonstrate that our relations hold for free theories.

\vskip 0.1in
\noindent
{\it\bf{Trace Anomalies and Boundary Central Charges:}}
We are interested in a classically Weyl-invariant theory embedded in a curved spacetime 
with a smooth and compact codimension-one boundary. 
The quantization requires regularization which results in a 
non-vanishing expectation value of the stress tensor trace.
The trace anomaly in a compact spacetime 
is well-known \cite{Deser:1993yx}. In particular there is no anomaly in odd dimensions. 
In the presence of a boundary, however, there are anomalies localized on the 
boundary, in both odd and even dimensions. 
These new anomalies have rich geometric stucture and they introduce new 
central charges that could be used to characterize the theories. 

Define the induced metric on the boundary as $h_{\mu\nu}= g_{\mu\nu}-n_\mu n_\nu$, where 
$n_{\mu}$ is an outward-pointing normal vector. 
The extrinsic curvature is $K_{\mu\nu}= h_\m^\lambda h_\n^\sigma \nabla_{\lambda} n_{\sigma}$ 
where $ \nabla_{\lambda}$ is the bulk covariant derivative.
We denote the traceless part of the extrinsic curvature as 
$\hat{K}_{\mu\nu}=K_{\mu\nu}-\frac{h_{\mu\nu}}{d-1}K$, which transforms 
covariantly under the Weyl transformation, and it plays an important role 
in constructing boundary Weyl invariants.

In $d=3$, the anomaly only appears on the boundary, and it is given by  
\cite{Graham:1999pm} 
\be
\langle T^{\mu}{}_{\mu}\rangle^{d=3} 
= \frac{\delta(x_{\perp})}{4 \pi}\left( a_{(3d)} \oR + b \tr \hat K^2\right)\ ,
\ee  
where  $\delta(x_{\perp})$ is a Dirac delta function with support on the boundary, and
$\tr \hat K^2= \tr K^2 - {1\over 2} K^2$; $\oR $ is the boundary Ricci scalar. 
For free fields, the values of these boundary charges were computed in the literature 
\cite{Nozaki:2012qd, Jensen:2015swa,Fursaev:2016inw}:
$a_{(3d)}^{s=0, (D)}=-{1\over 96}$,  $a_{(3d)}^{s=0, (R)}={1\over 96}$ 
and $a_{(3d)}^{s={1\over 2}}=0$, where $(D)/(R)$ 
denotes Dirichlet/Robin boundary condition. 
(In our notation, $s$ is the spin of the free field.)  
The $a_{(3d)}$ coefficient has been argued to decrease under 
boundary renormalization group flow~\cite{Jensen:2015swa}.

The structure becomes much richer in $d=4$ CFTs. 
The complete classification was recently given in \cite{Herzog:2015ioa}. 
Dropping a regularization dependent term, the trace anomaly reads
\be
\label{4dTrace}
&&\langle {T^\mu}_\mu \rangle^{d=4}=
{1\over 16 \pi^2} \Big( c W_{\mu\nu\lambda\rho}^2- a_{(4d)} E_4\Big)\\
\nn
&&+{\delta(x_{\perp})\over 16 \pi^2} \Big(a_{(4d)} E^{\rm{(bry)}}_4-b_1 \tr\hat{K}^3-b_2
h^{\alpha\gamma}\hat{K}^{\beta\delta}W_{\alpha\beta\gamma\delta} \Big) \ ,
\ee
where $E_4$ is the bulk Euler density in $d=4$, and $W_{\mu\nu\rho\sigma}$ is the Weyl tensor.
In the presence of a boundary, the 
boundary term of the Euler characteristic, $E^{\rm{(bry)}}$, is 
added in order to preserve the topological invariance.  
We refer readers to the literature for the values of the $a_{(4d)}$ and $c$ charges; these 
are the familiar central charges characterizing theories on a compact manifold. 
Let us list the values of the $b_1$ charge for free fields:
$b_1^{s=0,(D)}={2\over 35}$ \cite{JM}, 
$b_1^{s=0,(R)}={2\over 45}$  \cite{Moss}, 
$b_1^{s={1\over 2}}={2\over 7}$ \cite{Fursaev:2016inw}, 
$b_1^{s={1}}={16\over 35}$ \cite{Fursaev:2016inw}. 

Refs.\ \cite{Fursaev:2015wpa,Solodukhin:2015eca} observed that a relation $b_2=8c$ is universal 
for free theories. Such a relation can be better understood by studying the stress tensor two-point 
function carefully with a boundary. Two of us have argued  \cite{Herzog:2017xha} that the relation 
need not hold once interactions are included. For a theory with a line of fixed points, parametrized 
by $g$, we found that, 
perturbatively in the coupling, $b_2=8c+ {\cal O}(g^2)$.
Note that Wess-Zumino consistency implies that $a$-type central charges
cannot depend on marginal couplings \cite{OS:1991}.

The motivation of this letter is to generalize ref.\ \cite{Herzog:2017xha} to
consider other boundary charges in $d=3$ and $d=4$ CFTs.
It turns out that the general strategy is similar: one 
simply looks at the correlation functions of the displacement operator in flat space. 
But there are several differences when compared with the computation of the $b_2$ charge.
The first difference is that these $b$ and $b_1$ boundary charges do not talk to bulk 
charges, while the $b_2$ structure is intimately related to the surface term generated from 
varying the bulk $c$-type anomaly effective action. 
The second difference is that in order to compute $b_1$ in $d=4$, one has to look
not at two-point functions but at a boundary three-point function. 
We will conjecture a relation for $a_{(3d)}$, in terms of a 
boundary limit of the two-point function of the stress-tensor.

\vskip 0.1in
\noindent
{\it\bf{Displacement Operator and General Relations:}}
To set notation, let $W$ be the generating functional for connected Green's functions.  
The stress tensor in Euclidean signature is
\be
\la T_{\mu\nu} (x) \ra=-{2\over \sqrt{g}} {\delta W\over g^{\mu\nu}(x)} \ .
\ee 
 
Let us first consider $d=3$ CFTs with a boundary. 
Denote $\widetilde W$ as the anomalous part of $W$.
The anomaly effective action in dimensional regularization is 
\be
\widetilde W = \frac{\mu^\epsilon}{\epsilon} \frac{1}{4\pi} 
\left( a_{(3d)} \int_{\partial M} \oR + b \int_{\partial M} \tr \hat K^2 \right) \ .
\ee 
Consider the special case where $\partial M$ is almost the planar surface at $y=0$,
and can be described by a small displacement $\delta y(x^A)$,
which is a function of the directions tangent to the boundary, denoted by $x^A$.   
In this situation, the normal vector is well-approximated by
\be
n_\mu = (\partial_A \delta y, 1) \ .
\ee
The extrinsic curvature then becomes $K_{AB} = \partial_A \partial_B \delta y$,
and we have
\be
\label{eq1}
\int_{\partial M} \tr \hat K^2 = \frac{1}{2} \int_{\partial M}  \delta y \oBox^2 \delta y \ ,
\ee 
where $\oBox^2= \partial^A \partial_A$ acts only on the boundary.

Correlation functions of the displacement operator $D^n({\bf x})$ 
can be generated by varying $W$ with respect to $\delta y(x^A)$.  
Note that diffeomorphisms act on both the metric and the embedding function
$\delta y(x^A)$.  
As the effective action $W$ is diffeomorphism invariant, there is a Ward identity that
relates the stress tensor to the displacement operator, an 
integrated version of which in the flat limit becomes
\be
T^{nn}|_{\partial M}=D^n \ .
\ee    
Because the displacement operator lives inside the boundary surface, and because we have
conformal symmetry in this surface, the two point function 
is fixed up to a constant, which we call $c_{nn}$:
\be
\langle D^n({\bf x}) D^n(0) \rangle = \frac{c_{nn}}{{\bf x}^{2d}} \ .
\label{Dntwopointraw}
\ee  
(In the notation of \cite{Herzog:2017xha}, $c_{nn}$ was called $\alpha(1)$ 
through its relation to the two point function of the stress tensor.)
Replacing the expression (\ref{Dntwopointraw}) with a regularized version \cite{dimreg, Osborn:1993cr}
in the case of interest $d=3$,
\be
\langle D^n({\bf x}) D^n(0) \rangle 
= \frac{c_{nn}^{(3d)}}{512} \oBox^3 (\log \mu^2 {\bf x}^2)^2 \ ,
\ee
the scale-dependent part is then
\be
\label{eq2}
\mu {\partial\over \partial \mu} \langle D^n({\bf x}) D^n(0) \rangle 
=  \pi \frac{c_{nn}^{(3d)}}{32} \oBox^2 \delta({\bf x}) \ .
\ee 
Equating the scale dependent pieces yields
\be
\label{brel}
b = {\pi^2\over 8} c_{nn}^{(3d)} \ .
\ee
This relation was conjectured in \cite{Herzog:2017xha}, based on free theories 
\cite{Jensen:2015swa, Fursaev:2016inw}.
Here we have provided a general derivation.
A similar calculation for the case of a codimension-two defect in four-dimensions was presented in 
ref.\ \cite{Bianchi:2015liz} in the context of entanglement entropy.
Note that the $b$-charge can change under marginal 
deformations, although here we do not discuss a 3d example.

Next we consider $d=4$.
The constraint on the $b_2$ boundary charge was found in 
\cite{Herzog:2017xha}, and it reads
\be
\label{b2rel}
b_2= {2\pi^4\over 15} c^{(4d)}_{nn} \ .
\ee 
In flat space, the two-point function is 
not enough to constrain the $b_1$ boundary charge, since
the related Weyl anomaly has a ${\cal O}(K^3)$ structure. 
Thus, we will need to consider the three-point function.

The relevant anomaly effective action is
\be
\widetilde W^{(b_1)} 
= \frac{b_1}{16 \pi^2} \frac{\mu^\epsilon}{\epsilon} \int_{\partial M} \tr \hat K^3 \ . 
\ee
We again consider $\partial M$ to be nearly flat and described by a small 
displacement, $\delta y(x^A)$.
Approximating the normal vector by
$n_\mu = (\partial_A \delta y, 1)$, we obtain
\begin{align}
\begin{split}
\label{dispvariation}
\int_{\partial M} \tr \hat{K}^3 &  = \int_{\partial M} \Big( \tr\big[(\partial_A \partial_B \delta y)^3\big]   \\
& - (\oBox \delta y) \tr\big[(\partial_A \partial_B \delta y)^2\big]  + \frac{2}{9} (\oBox \delta y)^3 \Big)  \ .
\end{split}
\end{align}
We will relate this $b$-charge with the displacement operator three-point function defined by
\be
\label{3point}
\langle D^n ({\bf x}) D^n({\bf x'}) D^n({\bf 0}) \rangle 
= \frac{c_{nnn}}{|{\bf x}|^4 |{\bf x'}|^4 |{\bf x} - {\bf x}'|^4} \ ,
\ee 
where $c_{nnn}$ is a constant.  
The full structure of the stress tensor three-point function 
with a boundary has not been studied yet. 
But, as mentioned earlier, to constrain these
boundary charges one can simply look at the purely normal-normal component of the 
stress-tensor correlation functions that represent the displacement operator contributions.

While it is not obvious how to proceed in position space, we note that
 the Fourier transform 
of the three-point function of operators $O_1$, $O_2$ and $O_3$ 
is generally \cite{Bzowski:2013sza, Bzowski:2015pba}
\be
C_{123} \int_0^\infty d x \, x^\alpha \prod_{j=1}^3 p_j^{\beta_j} K_{\beta_j}(p_j x) \ ,
\ee
where $K_{\beta_j}(x)$ denotes the modified Bessel function of the second
kind, and
$\alpha = \frac{\delta}{2}-1, \beta_j = \Delta_j - \frac{\delta}{2}$; 
$\Delta_j$ is the conformal dimension of operator $O_j$
and $\delta$ is the dimension of the CFT.
In this case, we are interested in the CFT 
living on the boundary, so $\delta=3$ while the scaling dimension
of the displacement operator is $\Delta_j = 4$.
Taking $c_{123}$ as the corresponding coefficient of the 
position space three-point function,  one has
\cite{Bzowski:2013sza}
%
\be
c_{nnn} = \frac{105}{\sqrt{2} \pi^{5/2}} C_{nnn}\ .
\ee
The $1/x$ term in a small $x$ expansion of the integrand will give rise to a logarithm in the
position space three-point function and a corresponding anomalous scale dependence.
Observe that the $1/x$ term is
\begin{align}
\begin{split}
&\frac{3 \pi^{3/2}}{32 \sqrt{2}x} \Big(
p_1^6 + p_2^6 + p_3^6 - p_1^2 p_2^4 - p_1^2 p_3^4 
- p_2^2 p_1^4 
\\
&\hspace{0.8in}- p_2^2 p_3^4- p_3^2 p_1^4 - p_3^2 p_2^4 
- \frac{2}{3} p_1^2 p_2^2 p_3^2 \Big) \ .
\end{split}
\end{align}
Through integration by parts along the boundary, the 
above expression can be rewritten as
\begin{align}
\begin{split}
\frac{9\pi^{3/2}}{4 \sqrt{2} x} \Big(& (p_1 \cdot p_2)  (p_2 \cdot p_3)  (p_3 \cdot p_1) 
 \\
&\quad- p_1^2 (p_2 \cdot p_3)^2  + \frac{2}{9} p_1^2 p_2^2 p_3^2 \Big) \ . 
\end{split}
\end{align}
The result matches exactly the derivative form  \eqref{dispvariation} 
computed from the $b_1$ boundary trace anomaly.
Including a factor ${1\over 3!}$ coming from varying 
with respect to $\delta y$ three times, we obtain 
$b_1 = \frac{1}{3!} \cdot 16 \pi^2 \Big( \frac{9 \pi^{3/2}}{4 \sqrt{2}} \Big)
\Big( \frac{\sqrt{2} \pi^{5/2}}{105} \Big) c_{nnn}$, which gives
\be
\label{b1rel}
b_1 
&=& \frac{2 \pi^6}{35} c_{nnn} \ .
\ee  
This boundary charge in $d=4$ can depend on marginal interactions.
In particular, if the charge $b_2$ of the mixed-dimensional quantum 
electrodynamics (QED) depends on the marginal 
interactions \cite{Herzog:2017xha}, so does $b_1$.

\vskip 0.1in
\noindent
{\it\bf{Conjecture for $a_{(3d)}$:}}
From refs.\ \cite{McAvity:1993ue,McAvity:1995zd,Herzog:2017xha}, we can write 
down expressions for the near-boundary limit of the stress-tensor two-point function:
\be
\langle T_{\mu\nu}({\bf x},y) T_{\rho \sigma}({\bf 0},y') \rangle 
= A_{\mu\nu,\rho\sigma}({\bf x},y,y') \frac{1}{|{\bf x}|^{2d} }\ ,
\ee
where
\begin{align}
\begin{split}
A_{nn,nn}({\bf x},y,y') &= \alpha(v)\,,
\\
A_{nA,nB}({\bf x},y,y') &= -\gamma(v) I_{AB}({\bf x},y,y')  ,
\\
A_{AB,CD}({\bf x},y,y') &= 
\alpha(v)\frac{d}{d-1}I_{AB,CD}^{(d)} \\
 +& \left(2 \epsilon(v)  - \frac{d}{d-1}\alpha(v)  \right) I_{AB,CD}^{(d-1)}  \ ,
\end{split}
\end{align}
where $I_{AB}(x) = \delta_{AB} - 2\frac{x_A x_B }{x^2}$ and $ I^{(d)}_{AB,CD} 
= \frac{1}{2}(I_{AC}I_{BD} + I_{AD}I_{BC}) - \frac{1}{d} \delta_{AB} \delta_{CD}$. 
The quantity $v$ is a cross-ratio $v = \frac{(x-x')^2}{(x-x')^2 + 4 y y'}$, which behaves 
as $\sim 1 - \frac{4yy'}{|{\bf x}|^2}$ near the boundary at $v=1$. 

The functions $\alpha$, $\gamma$ and $\epsilon$ are related to each other by two differential constraints. 
Conservation of the stress tensor at the boundary, conformal invariance, and 
unitarity together impose that $\gamma$ smoothly 
vanishes as $v\to 1$, while $\alpha$ is smooth, and $\epsilon $ can blow up as $(1-v)^{\delta-1}$ for 
a small anomalous dimension $\delta >0$. 
Both $\alpha$ and $\epsilon$ may have $O(1-v)^0$ terms, which we refer to as $\alpha(1)$ and $\epsilon(1)$. 
(Note the relation between $\alpha(v)$ and the $D^n$ two-point function, $\alpha(1) = c_{nn}$.) 

The symmetries also allow for a distributional term in the two-point function $C I_{AB,CD}^{(d-1)} \delta(y)\delta(y')$. 
This term, if present, indicates a conserved stress tensor purely on the boundary, as would arise from decoupled boundary degrees of freedom.

We conjecture that the boundary anomaly coefficient $a_{(3d)}$ is a linear combination of $\alpha(1)$, $\epsilon(1)$, and $C$. 
The dependence on $C$ is already fixed by the argument relating the trace anomaly of a two-dimensional CFT to the two-point function of its stress tensor. 
More precisely, $c_{(2d)}= 2\pi C$, where $c_{(2d)}$ is the 2d central charge in the Euler anomaly $\la T^A_A \ra = \delta(y) \frac{c_{(2d)}}{24\pi} \oR$.
We fix the dependence on $\alpha(1)$ and $\epsilon(1)$ by the known values for the conformal scalar with Dirichlet and Robin boundary conditions, giving
\be
\label{arel}
a_{(3d)} =\frac{\pi^2}{9} \left( \epsilon(1) -  \frac{3}{4}\alpha(1)+3 C \right)  \ ,
\ee 
where $C$ vanishes for a theory of free $3d$ scalars and for free $3d$ fermions.
Note this conjecture gives the correct result for free fermions, reproducing 
$a_{(3d)}^{s=\frac{1}{2}}=0$.

In a general interacting bCFT we suspect only $\alpha(1)$ to be nonzero for the following reason. 
Interactions coupling boundary degrees of freedom to the bulk ought to lead to a unique stress tensor, leading to $C=0$. 
Meanwhile, $\epsilon(1)$ corresponds to a dimension$-3$ boundary operator appearing 
in the boundary operator product expansion of $T_{AB}$, but the boundary conformal symmetry 
does not guarantee the existence of such an operator.

Reflection positivity means that the functions $\alpha(v)$ and $\epsilon(v)$ are non-negative \cite{Herzog:2017xha}. 
The coefficient $C$ is also non-negative. 
If $\epsilon(v)$ is regular near the boundary, then $\epsilon(1)$ is non-negative, and 
comparing with the new result~\eqref{brel} for $b$, we obtain the bounds
\be
\label{bound3d} 
{\rm d=3~~bCFTs}: \quad \frac{a_{(3d)}}{b}\geq -\frac{2}{3} \,,~~~  (b\geq 0)\,.
\ee
These bounds recall the Hofman-Maldacena \cite{Hofman:2008ar} bounds on $d=4$ bulk central charges. 
However, if $\epsilon(v)$ is singular near the boundary, then there is no constraint on the sign of $\epsilon(1)$, and 
thus, no definite bound on $a_{(3d)}$ charge.

\vskip 0.1in
\noindent
{\it\bf{Two- and Three-Point Functions in Free Theories:}}
We would like to verify the general relations \eqref{brel} and \eqref{b1rel} in free 
theories, including a conformal scalar, a Dirac fermion and, in $d=4$, Maxwell theory.

The stress tensor two-point functions with a planar boundary for the 
scalar and fermion were already considered in ref.\ \cite{McAvity:1993ue}.
More recently, ref.\  \cite{Herzog:2017xha} computed the two-point functions
for a Maxwell field.
We will list the relevant two-point function results for completeness,  
and consider three-point functions with a boundary in free theories.  
These latter results are, to our knowledge, new. 

Considering first a vector of scalar fields, i.e $\phi \to \phi^a$ 
(the index $a$ will be suppressed), we 
introduce complementary projectors $\Pi_\pm$ satisfying 
$\Pi_+ + \Pi_- = \id$ and $\Pi_\pm^2 = \Pi_\pm$. 
The boundary conditions are $\partial_n  (\Pi_+ \phi)|_{y=0} = 0$ and $\Pi_- \phi|_{y=0} = 0$.
The scalar displacement operator is 
\be
T_{nn} = (\partial_n \phi)^2 - \frac{1}{4} \frac{1}{d-1} \left( (d-2) \partial_n^2 + \Box \right) \phi^2 \ ,
\ee 
which is the boundary limit of the normal-normal component of the improved stress tensor.
The two-point function of the scalar field can be found using the image method:
\begin{align}
\begin{split}
\langle \phi(x) \phi(x') \rangle = &\K \Big( \frac{\id}{|x-x'|^{d-2}}
\\
&+ \frac{\chi}{(({\bf x} - {\bf x}')^2 + (y+y')^2 )^{(d-2)/2}} \Big) \ ,
\end{split}
\end{align} 
where the parameter $\chi=\Pi_+-\Pi_-$ is determined by boundary conditions. 
We have adopted the normalization 
$\K= {1\over (d-2) {\rm{Vol}}(S^{d-1})}$ where ${\rm Vol}(S^{d-1})
= {2 \pi^{d\over 2}\over \Gamma{({d\over 2})}}$.
Note $\chi^2 = \id$, and that an eigenvalue of $\chi$ is $1$ for Neumann 
and -1 for Dirichlet boundary conditions.

To keep the expressions simple, we will focus on the displacement operator 
two-point function in $d=3$ and the three-point function in $d=4$. 
These two quantities are required in computing the boundary 
central charges from the relations \eqref{brel} and \eqref{b1rel}.  

A straightforward application of Wick's theorem gives 
\begin{align}
\label{3d2pointphi}
\langle  D^n({\bf x}) D^n({\bf 0})   \rangle^{s=0}_{3d} &
 = { \tr(\id) \over 8\pi^2 {\bf x}^{6} } \ , \\
\langle D^n({\bf x}) D^n({\bf x'})   D^n({\bf 0}) \rangle^{s=0}_{4d} &
=  \nonumber \\
 &
{1 \over 9 \pi^6} {8 \tr( \id)-  \tr(\chi) \over  |{\bf x}|^4 |{\bf x'}|^4 |{\bf x} - {\bf x'}|^4}
\label{4d3pointphi}
\ .
\end{align} 
The result \eqref{3d2pointphi} implies that the $b$ boundary charge 
(in $d=3$)
does not depend on boundary conditions for a free scalar. 
Indeed, using the relation \eqref{b1rel}, we recover the known value 
of the $b$ charge for a $d=3$ free scalar,  $b={1\over 64}$.
On the other hand, clearly $b_1$ is sensitive to boundary conditions through
the $\tr(\chi)$.
Using the relation \eqref{b1rel}, we can verify that $b_1$ is
 ${2\over 35}$ for a Dirichlet scalar 
and ${2\over 45}$ for a Neumann scalar.

Next we consider a Dirac fermion. 
In Minkowski (mostly plus) signature, $\{ \gamma_\mu, \gamma_\nu \} = - 2 \eta_{\mu\nu}$.  
The fermion's displacement operator and two-point function are 
\be
T_{nn} &=& \frac{i}{2} \left( \dot{\bar \psi} \gamma_n \psi 
- \bar \psi \gamma_n \dot \psi \right)\ , ~~ (\dot \psi \equiv \partial_n \psi)\\
\langle \psi(x) \bar \psi(x') \rangle &=& 
 \K_f \Big( \frac{i \gamma \cdot (x - x')}{|x-x'|^d} 
+ \chi {i \gamma \cdot (\bar x - x')\over |\bar x - x'|^d} \Big) \ ,
\ee
where $\bar x = (-y,{\bf x})$ and $\K_f =1 / \Vol(S^{d-1})$ 
and $\bar \psi = \psi^\dagger \gamma^0$. 
The $\chi$ parameter satisfies 
\be
\chi \gamma_n = - \gamma_n \bar \chi ~,~~ \chi \gamma_A 
= \gamma_A \bar \chi~, ~~\chi^2 = \bar \chi^2 = \id \ ,
\ee   
where $\bar \chi= \gamma^0 \chi^\dagger \gamma^0$. 
Focusing on the fermion displacement operator two-point function in $d=3$ 
and the three-point function in $d=4$, we find  
\be
\label{3d2pointpsi}
\la D({\bf x}) D({\bf 0}) \ra^{s={1\over 2}}_{3d} 
&=& \frac{3}{16 \pi^2} {\tr_\gamma (\id) \over{\bf x}^{6}}\ , \\
\label{4d3pointpsi}
\la D({\bf x}) D({\bf x'}) D({\bf 0}) \ra^{s={1\over 2}}_{4d}
&=&\frac{5}{4 \pi ^6 }{\tr_\gamma (\id)\over {\bf x}^4 {\bf x'}^4 ({\bf x}-{\bf x'})^4}\ ,
\ee  
where $\tr_\gamma (\id)$ depends on the Clifford algebra 
one uses; we will take $\tr_\gamma(\id) = 2^{\lfloor d/2 \rfloor}$. 
As $\chi^2 = \id$, the boundary dependence drops out of these
two- and three-point functions.   
We can again verify the relations \eqref{brel} and \eqref{b1rel} for the fermion.

Finally, we consider a Maxwell field in Feynman gauge. 
As the field in $d=3$ is not conformal, we focus on 
the $d=4$ case. 
The displacement operator is 
\be
T_{nn} = \frac{1}{2} F_{nA} {F_n}^A - \frac{1}{4} F_{AB} F^{AB} \ , 
\ee 
and the gauge field two-point function is
\begin{align}
\begin{split}
\la A_{\mu}(x) A^{\nu}(x') \ra
=\K& \Big( \frac{\delta^{\nu }_{\mu}}{(x-x')^{2}}
\\
&+ \frac{\chi^\nu_\mu}{(({\bf x}-{\bf x}')^2 + (y+y')^2)^{2}} \Big)\,.
\end{split}
\end{align} 
The $\chi^\nu_\mu$ parameter determines the boundary condition;
it is equal to $\delta^\nu_\mu$ up to a sign.
For gauge fields one can consider the absolute boundary condition 
where the normal component of the field strength 
is zero, which gives $\partial_n A_{A} = 0$ and $A_{n} = 0$, or 
the relative boundary condition where $A_{A} = 0$ which gives $\partial_n A^{n} = 0$ 
when recalling the gauge fixing. See ref.\ \cite{Herzog:2017xha} for more details. 
We find
\be
\langle D^n({\bf x}) D^n({\bf x}') D^n({\bf 0})\rangle^{s=1}_{4d} 
&=&  \frac{512 \kappa^3}{|{\bf x}|^4 |{\bf x}'|^4 |{\bf x} - {\bf x}'|^4} \ ,
\ee 
independent of the choice of boundary conditions.
From the relation \eqref{b1rel} 
we recover the value of $b_1$ charge for the $d=4$ Maxwell field with a boundary.

\vskip 0.1in
\noindent
{\it\bf{Discussion:}} 
We presented new results for the boundary terms in the trace anomaly
for CFTs in 3d and 4d.  
By relating $b$ (\ref{brel}), $b_1$ (\ref{b1rel}), $b_2$ (\ref{b2rel}), and $a_{(3d)}$ (\ref{arel})
to two- and three-point functions of the displacement operator and stress tensor
 in flat space, these results make the boundary coefficients more straightforward to compute. 

While we proved the relations (\ref{brel}) and (\ref{b1rel}) in this letter, 
two of us demonstrated (\ref{b1rel}) previously \cite{Herzog:2017xha}, and 
(\ref{arel}) remains a conjecture along with the lower bound  (\ref{bound3d})
that follows from it (with the caveat discussed there).
Ultimately, perhaps building on the bound (\ref{bound3d}), 
we hope that a classification scheme for bCFT can be organized around these coefficients.
%
We suspect bounds on the 4d coefficients $b_1$ and $b_2$ exist as well, 
beyond $b_2\geq 0$  \cite{Herzog:2017xha}.

Finally, extending the 3d results to the case of a 4d bulk and 2d defect, there are
applications of these results to quantum entanglement (see ref.\ \cite{Bianchi:2015liz}
for results along these lines). 

\vskip 0.1in
\noindent
{\bf Acknowledgements:} 
We thank D.~Gaiotto for useful discussions. 
The work of C.P.H. and K.-W.H. was supported in part by the 
National Science Foundation under Grant No.\ PHY-1620628. 
The work of K.J. was supported in part by the US 
Department of Energy under Grant No. DE-SC0013682.

\end{document}